\documentstyle[pra,aps,epsf]{revtex}

\def \lsim {\,{\scriptscriptstyle{\stackrel{<}{\sim}}}\,}

\newcommand{\nn}{\noindent}
\newcommand{\Ref}[1]{(\ref{#1})}
\newcommand{\beq}{\begin{equation}}
\newcommand{\eeq}{\end{equation}}
\newcommand{\bea}{\begin{eqnarray}}
\newcommand{\eea}{\end{eqnarray}}
\newcommand{\beao}{\begin{eqnarray*}}
\newcommand{\eeao}{\end{eqnarray*}}
\newcommand{\m}{{\,\mbox{\rm m}}}
\newcommand{\mum}{{\,\mu{\mbox{\rm m}}}}

\begin{document}
\draft
\thispagestyle{empty}
\title{
Stronger constraints for nanometer scale Yukawa-type
hypothetical interactions from \\ the new measurement
of the Casimir force
}
\author{
M.~Bordag,\footnote{Electronic address: Michael.Bordag@itp.uni-leipzig.de}
B.~Geyer,\footnote{Electronic address: geyer@itp.uni-leipzig.de}
G.L.~Klimchitskaya,\footnote{on leave from North-West Polytechnical 
Institute, St.Petersburg, Russia.\\  
Electronic address:  galina@GK1372.spb.edu}${}^{,}$\footnotemark[5]
 and V.M.~Mostepanenko\footnote{on leave from A.Friedmann Laboratory
for Theoretical  Physics, St.Petersburg, Russia.\\ 
Electronic  address: mostep@fisica.ufpb.br
}${}^{,}$\footnote[5]{Present address: Department of Physics, 
Federal University
of Paraiba, C.P. 5008, \\ CEP 58059-970, Joao Pessoa, Pb-Brazil}
}

\address
{Institute for Theoretical Physics, Leipzig
University,  Augustusplatz 10/11, 04109 Leipzig, 
Germany}
\maketitle
\begin{abstract}
We consider the Casimir force including all important corrections
to it for the configuration used in a recent experiment
employing an atomic force microscope. We calculate the long-range
hypothetical forces due to the exchange of light and massless
elementary particles between the atoms constituting the bodies
used in the experiment --- a dielectric plate and a sphere both
covered by two thin metallic layers.
The corrections to these forces caused by the small surface 
distortions were found to be essential for nanometer Compton wave
lengthes of hypothetical particles. 
New constraints for the constants of Yukawa-type interactions are
obtained from the fact that such interactions were not observed 
within the limits of experimental accuracy. They are stronger
up to 140 times in some range than the best constraints known
up date. Different possibilities are also discussed to strengthen
the obtained constraints in several times without principal changes
of the experimental setup.
\end{abstract}


%%%%%%%%%%%%%%%%%%%%%%%%%%%%%%%%%%%%%%%%%%%%%%%%%%%%%%%%%%%%%%%%% %%%%%

\section{INTRODUCTION}

Hypothetical interactions of Yukawa- and power-types
have been discussed  actively in recent years (for a
collection of references on this subject, see [1]).
They may be considered as specific corrections to the
classical gravitational force at small distances [2].
An alternative interpretation comes from the elementary
particle physics. According to the unified gauge
theories, supersymmetry and supergravity [3], there
would exist a number of light or massless elementary
particles (for example, axion, scalar axion,
graviphoton, dilaton, arion, and others). The
exchange of such particles between two atoms gives 
rise to an interatomic force described by Yukawa
or power-law effective potentials. The interaction range
of this force is to be considered from one angstr\"{o}m to
hundreds of kilometers. Because of this, it is called
a ``long-range force" (in comparison with the
nuclear size).

The constraints for the constants of these hypothetical
long-range interactions lead to new knowledge about
the parameters of associated elementary particles.
Such constraints are obtainable from Galileo-,
E\"{o}tv\"{o}s- and Cavendish-type experiments, from
the measurements of the van der Waals and Casimir
force, transition probabilities in exotic atoms,
etc [4]. The pioneering studies in the application of the
Casimir force measurements to the problem of long-range
interactions were made in Refs.\ [5--7]. There it was shown 
that the Casimir effect
leads to the strongest constraints on the constants
of Yukawa-type interactions with a range of action
of $10^{-8}\m\,<\lambda < 10^{-4}\m$ 
[4,5,7].

In the beginning of 1997 the results 
 on the demonstration of the Casimir force between
the metallized surfaces of a flat disc and a spherical
lens were published [8]. 
The absolute error of the force measurements was
$\Delta F\approx 10^{-11}\,$N for distances
between the disc and the lens from one to six
micrometers. This corresponds to the relative error
of approximately 5\% at the smallest surface
separation. Some tentative conclusions from
experimental observations of [8] were made in [9,10]
concerning the possible constraints on long-range
interactions. The detailed and accurate analyses of
constraints following from [8] was given in [11]
for both Yukawa- and power-type interactions with
account of all different corrections to the Casimir
force. It was established that the constraints for
Yukawa-type interactions following from [8] are the
best ones within a range
$2.2\times 10^{-7}\m \leq\lambda\leq 1.6\times10^{-4}
\m$. The fact was emphasized that they surpass 
the previously known constraints in this range up to
factor of 30.

Recently, the results of a new experiment were 
published [12] on the precision measurement of 
the Casimir force between a metallized sphere and
a flat plate. The  force was measured using an atomic
force microscope for the plate-sphere separations
from 120\,nm to 900\,nm.  In [12] the
corrections to the Casimir force were taken 
into account due to the finite conductivity of the
metal and due to the small surface distortions
(the temperature corrections are negligible in the
range of the measurement). The theoretical value of the
Casimir force including corrections was confirmed
experimentally with the root mean square deviation
$\sigma=1.6\times 10^{-12}\,$N. This gives the
possibility to use the experiment [12] for obtaining
much stronger constraints on the hypothetical long-range
interactions at the nanometer scale.

In this paper we calculate the hypothetical forces which 
might appear between a sphere and a flat plate.
It is shown that the surface distortions contribute
to the value of the hypothetical force essentially
at the nanometer scale and influence the strength of
the resulting constraints. As indicated below, the
experiment [12] imposes the strongest constraints
for Yukawa hypothetical interactions within a range
$5.9\,$nm$ \leq\lambda\leq 100\,$nm.
They are stronger than the previously known
constraints in this range up to a factor of 140.
We notice that both recent experiments on Casimir
force measuring [8,12] lead to stronger constraints
for Yukawa-type hypothetical interactions, and yet these
constraints hold for different $\lambda$-ranges
which do not intersect each other.

The paper is organized as follows. In Sec.~II the
necessary details of the experiment [12] are
summarized and the theoretical results for the
Casimir force with all the necessary corrections are
discussed. Section III is devoted to the calculation
of the hypothetical forces in the experimental
configuration of [12]. Both the layer structure of the
test bodies and small surface distortions are taken
into account. Section IV contains the derivation of
the new constraints for the parameters of Yukawa-type
hypothetical interactions following from the
experiment [12]. The possible improvement of the
experimental scheme is also discussed in order to
provide  stronger constraints. Section V contains
the conclusions and discussions.

\section{THE CASIMIR FORCE BETWEEN A SPHERE AND
A DISC}

In the experiment [12] the Casimir force was measured
between a metallized polystyrene sphere of 
radius $R=98\mum$ and a sapphire disc of 
diameter $D=1.25\,$cm (see Fig.~1).
The sphere was attached to a cantilever of an atomic
force microscope. The sphere and the disc were
covered by a layer of $Al$ with
$\Delta_1^{\prime}=\Delta_1=300\,$nm thickness. Both
surfaces were covered then by a layer of
$60\%Au/40\%Pd$ with
$\Delta_2^{\prime}=\Delta_2=20\,$nm (the thicknesses
in Fig~1 are shown to be different on both bodies for
generality). Disc-sphere surface separations lie in
the range 
$120\,$nm$\leq\lambda\leq 900\,$nm.

The Casimir force in the configuration of Fig.\ 1 is
the same as in the configuration of a spherical lens
above an infinite disc due to inequalities
$a\ll R\ll D$:
\begin{equation}
F^{(0)}(a)=-\frac{\pi^3}{360}\,R\,\frac{\hbar c}{a^3}.
\label{1}
\end{equation}
\nn
This formula was derived in [13] for the first time and
reobtained by different authors afterwards (see, e.g.,
[14--16]).

In the range of $a$ under consideration the substantial
corrections to Eq.\ (1) are due to the finite
conductivity of the metal and due to the surface
distortions (the temperature corrections are important
for larger surface separations [11]). The corrections due
to the finite conductivity can be obtained by the 
use of perturbation theory in the small parameter
$\delta_0/a=c/(\omega_p a)=\lambda_p/(2\pi a)$,
where $\omega_p (\lambda_p)$ is the effective plasma
frequency (wave-length) of the electrons,
$\delta_0$ is the penetration depth of the 
electromagnetic oscillations into the metal. For
$Al$ we have 
$\lambda_p\approx 100\,$nm [12] (the external metallic
layer is rather transparent, so it is equivalent to several
nanometers of $Al$ only). 
Thus, the value of
$\delta_0/a$ changes from 0.133 to 0.018 in the
range of the measurement.

The first order correction to (1) was firstly found
in [17] for the configuration of two plane parallel
plates (see also [18]). The second order correction was
obtained in [19]. We finally obtain for the case of
plane parallel plates
\begin{equation}
F_{\delta_0}(a) \approx
 F^{(0)}(a)\left(1-\frac{16}{3}\,\frac{\delta_0}{a}+
24\,\frac{\delta_0^2}{a^2}\right).
\label{2}
\end{equation}

From the exact expression of $F_{\delta_0}$ (before
the expansion in powers of $\delta_0/a$) it is quite
clear that $F_{\delta_0}$ is sign-constant for all
$\delta_0$  and tends to zero in the formal limit
$\delta_0\to\infty$. This allows us to obtain a simple
interpolation formula. It gives the same result as
(2) for small $\delta_0/a$, but applicable over 
a broader interval
$0\leq\delta_0/a\lsim 0.2$ [14]
\begin{equation}
F_{\delta_0}(a) \approx
 F^{(0)}(a)\left(1+\frac{11}{3}\,\frac{\delta_0}{a}
\right)^{-16/11}.
\label{3}
\end{equation}

It is an easy matter to expand (3) in powers of
$\delta_0/a$ and to modify the result to the case
of a sphere above a disc by the use of
the force proximity theorem [15]. 
The result up to fourth order is
\beq
F_{\delta_0}(a)\equiv F^{(0)}(a)+\Delta_{\delta_0}F^{(0)}(a)
\approx F^{(0)}(a)\left(1-4\,\frac{\delta_0}{a}+
\frac{72}{5}\,\frac{\delta_0^2}{a^2}-
\frac{152}{3}\,\frac{\delta_0^3}{a^3}+
\frac{532}{3}\,\frac{\delta_0^4}{a^4}
\right).
\label{4}
\eeq

Let us now discuss the corrections to (1) due to
surface distortions. The character of distortions was
investigated in [12] using an atomic force 
microscope. They look like some crystal boxes on 
both surfaces with a mean height $h=35\,$nm.
These boxes are spaced rather rarely in a stochastic
way, so that the
ratio of the occupied surface area to the free one is
$\beta\approx 0.11$ (see [20] for the detailed investigation 
of the surface distortions 
 in [12] and their contribution 
to the Casimir force). The general form of the corrections
to Eq.\ (1) due to distortions was obtained in [21]
by the use of perturbation theory in the small parameters
$A_1/a$, $A_2/a$. Here $A_{1,2}$ are the distortion
amplitudes on both test bodies. They are defined in
such a way that the mean values of the distortion
functions $A_i f_i(x_i,y_i)$ of the first and  the
second body are equal to zero, and 
$\max |f_i|=1$. For two plane parallel plates the
result up to the fourth order inclusive is [21]

\bea
&&F_d(a)\equiv F^{(0)}(a)+\Delta_d F^{(0)}(a)
\approx F^{(0)}(a)\left[1+c_{pp}^{(1)}\left(
\langle f_1^2 \rangle \frac{A_1^2}{a^2}-
2\langle f_1 f_2 \rangle \frac{A_1 A_2}{a^2}+
\langle f_2^2 \rangle \frac{A_2^2}{a^2}\right)
\right. \nonumber \\
&&\phantom{aaaaaaa}
+c_{pp}^{(2)}\left(
\langle f_1^3 \rangle \frac{A_1^3}{a^3}-
3\langle f_1^2 f_2 \rangle \frac{A_1^2 A_2}{a^3}+
3\langle f_1 f_2^2 \rangle \frac{A_1 A_2^2}{a^3}-
\langle f_2^3 \rangle \frac{A_2^3}{a^3}\right)
 \label{5} \\
&&\phantom{aaaaaaaaaa}
\left.
+c_{pp}^{(3)}\left(
\langle f_1^4 \rangle \frac{A_1^4}{a^4}-
4\langle f_1^3 f_2 \rangle \frac{A_1^3 A_2}{a^4}+
6\langle f_1^2 f_2^2 \rangle \frac{A_1^2 A_2^2}{a^4}-
4\langle f_1 f_2^3 \rangle \frac{A_1 A_2^3}{a^4}+
\langle f_2^4 \rangle \frac{A_2^4}{a^4}\right)
\right]. \nonumber 
\eea

\nn
Here the angle brackets denote the averaging procedure over
the area of the plates and the coefficients are
$c_{pp}^{(1)}=10$, $c_{pp}^{(2)}=20$,
$c_{pp}^{(3)}=35$.

The result \Ref{5} may be modified for a configuration of 
a sphere above a disc by the use of
the force proximity theorem [15] which works good in the case
of stochastic distortions [22] 
(as was noticed in [22] for the case of large-scale deviations 
on the boundary surfaces from
the perfect shape the appropriate redefinition of the distance
between the interacting bodies is necessary for the correct
application of this theorem).
Again, it has the form of
\Ref{5} where the numerical coefficients
$c_{pp}^{(i)}$ should be changed for
$c_{ps}^{(1)}=6$, $c_{ps}^{(2)}=10$,
$c_{ps}^{(3)}=15$.

For the experiment [12] 
$A_1=A_2=h/(1+\beta)\approx 31.5\,$nm, so that the
expansion parameter in \Ref{5} changes from 0.26 to
0.035 (note that, contrary to \Ref{4}, the
expansion \Ref{5} starts from the second-order term).
The minimal distance between the tops of two distortions
situated against each other is equal to 50\,nm. It is
still in the action range of the Casimir forces.
The quantities of the form 
$\langle f_1^i f_2^k\rangle$ depend on the phase shifts 
$\varphi_x, \varphi_y$ 
between the distortions situated on different
bodies. The measured Casimir force was averaged in [12]
over 26 scans. Because of this it is necessary to
consider the quantity
$\langle\Delta_d F^{(0)}(a)
\rangle_{\varphi_x,\varphi_y}$
averaged over the possible values of phase shifts [20].
With a required accuracy we can use the sum of
corrections to (1) due to finite conductivity and
surface distortions. As a consequence of 
Eqs.\ \Ref{4}, \Ref{5}, the theoretical value of the
Casimir force is:
\beq
F(a)=F^{(0)}+\Delta_{\delta_0}F^{(0)}(a)+
\langle\Delta_d F^{(0)}(a)
\rangle_{\varphi_x,\varphi_y}.
\label{6}
\eeq
\nn

As it was shown in [12,\,20], the deviation of 
the theoretical value \Ref{6} from the experimental
results is less than the absolute error of force
measurements 
$\Delta F$ within the most
interesting range 
120\,nm$\leq a\leq 300\,$nm
(note, that in [12] there was approximately
$\Delta F\approx 2\times 10^{-12}\,$N [20]).
This fact is used in Sec.\ IV for obtaining stronger 
constraints for hypothetical long-range interactions.

\section{CALCULATION OF THE HYPOTHETICAL FORCES}

As noted above, for the experimental configuration
of [12] the disc can be considered as of infinite diameter. 
Let us start with Yukawa-type hypothetical interactions
and calculate firstly the force, acting between a
homogeneous disc and a sphere. The potential between
two atoms which are separated by a distance $r_{12}$
and belonging to different bodies is
\begin{equation}
V_{Yu}=-\alpha N_1N_2\hbar c\frac{1}{r_{12}}
e^{-r_{12}/\lambda},
\label{7}
\end{equation}
\noindent
where $\alpha$ is a dimensionless constant,
$\lambda=\hbar/(mc)$ is the Compton wavelength of some
hypothetical particle giving rise to new interaction,
$N_i$ are the numbers of nucleons in the
atomic nuclei. They are introduced into \Ref{7} to
take off the dependence of $\alpha$ on the sort of
atoms [23].

The potential energy of a sphere above a disc (plate)
is given by
\beq
U_{Yu}^{(sp)}(a)=-\alpha N_1 N_2 \hbar c
\int\limits_{V_s} d^3r_1
\int\limits_{V_p} d^3r_2
\frac{1}{r_{12}}e^{-r_{12}/\lambda}.
\label{8}
\eeq
\nn
For the wavelength $\lambda$ at nanometer scale the
plate may be considered not only as of infinite area
but as of infinite width also. Integrating in \Ref{8}
over $V_p$ one obtains
\beq
U_{Yu}^{(sp)}(a)=-2\pi\alpha N_1 N_2 
\lambda^2\hbar c
\int\limits_{V_s} d^3r_1
e^{-z_{1}/\lambda}.
\label{9}
\eeq
\nn
Here the $(x,y)$-plane of the coordinate system 
coincides with the surface of the plate and the $z$-axis is
perpendicular to it (see Fig.~1).

Integrating over $x_1,y_1$ in \Ref{9} and calculating
the force as derivative
\beq
F_{Yu}^{(sp)}(a)=-
\frac{\partial U_{Yu}^{(sp)}(a)}{\partial a},
\label{10}
\eeq
\nn
we come to the expression
\beq
F_{Yu}^{(sp)}(a)=-2\pi^2\alpha N_1 N_2 
\lambda\hbar c
\int\limits_{a}^{a+2R} dz_1
\left[R^2-(z_1-a-R)^2\right]
e^{-z_{1}/\lambda}.
\label{11}
\eeq

After the integration over $z_1$ the result is
\beq
F_{Yu}^{(sp)}(a)=-4\pi^2\alpha N_1 N_2 
\lambda^3\hbar c
e^{-a/\lambda}\,\Phi(R,\lambda),
\label{12}
\eeq
\nn
where the notation 
\beq
\Phi(R,\lambda)=R-\lambda+
(R+\lambda)\,e^{-2R/\lambda}
\label{13}
\eeq
\nn
is introduced.

Eq.~\Ref{12} may be rewritten in terms of 
the densities of the sphere
and the disc, respectively,
$\rho^{\prime}=1.06\times 10^3\,$kg/m${}^3$,
$\rho=4.0\times 10^3\,$kg/m${}^3$
\beq
F_{Yu}^{(sp)}(a)=-4\pi^2\alpha 
\frac{\hbar c}{m_p^2}
\lambda^3
e^{-a/\lambda}\,\Phi(R,\lambda),
\label{14}
\eeq
\nn
where $m_p$ is the proton mass.

Using \Ref{14} it is an easy task to calculate
the force acting in the configuration of the experiment [12]
(see Fig.~1), i.e., taking into account the metallic
layers on the sphere and the disc. For this purpose
the contribution of, e.g., a layer on the sphere, may
be represented as the difference of two quantities,
given by \Ref{14}, with the appropriate densities,
radii and distances to the disc. The complete
force acting in the configuration of Fig.~1 consists of
twenty five contributions of the form of \Ref{14}.
After some rearrangements the result is

\bea
&&F_{Yu}(a)=-4\pi^2\alpha 
\frac{\hbar c}{m_p^2}\rho \rho^{\prime}
\lambda^3
e^{-\frac{a}{\lambda}}\,
\left[\rho_2-(\rho_2-\rho_1)
e^{-\frac{\Delta_2}{\lambda}}-(\rho_1-\rho)
e^{-\frac{\Delta_2+\Delta_1}{\lambda}}\right]
\label{15} \\
&&
\phantom{aaaaaa}
\times
\left[\rho_2^{\prime} \Phi(R,\lambda)-
(\rho_2^{\prime}-\rho_1^{\prime})\Phi(R_1,\lambda)
e^{-\frac{\Delta_2^{\prime}}{\lambda}}-
(\rho_1^{\prime}-\rho^{\prime})\Phi(R_2,\lambda)
e^{-\frac{\Delta_2^{\prime}+
\Delta_1^{\prime}}{\lambda}}
\right].
\nonumber
\eea
\nn
Here $\rho_2^{\prime},\, \rho_2$ are the densities
of the external layers on the sphere and the disc,
and
$\rho_1^{\prime},\, \rho_1$ are the internal ones.
Also the following notations are introduced:
$R_1=R-\Delta_2^{\prime}$,
$R_2=R-\Delta_1^{\prime}-\Delta_2^{\prime}$.

One may hope to get strong constraints on $\alpha$
within the range $\lambda\sim a$ only. With regard
to $a\ll R$ it follows $\lambda\ll R$ and
$\Phi(R,\lambda)\approx R$ for nanometer scale of  
$\lambda$ we are concerned with. The result \Ref{15}
may be simplified additionally when it is 
also taken into account that in the experiment [12]
$\rho_1^{\prime}=\rho_1=2.7\times 10^3\,$kg/m${}^3$
and
$\rho_2^{\prime}=\rho_2=16.2\times 10^3\,$kg/m${}^3$

\bea
&&F_{Yu}(a)=-4\pi^2\alpha 
\frac{\hbar c}{m_p^2}
\lambda^3
e^{-\frac{a}{\lambda}}\,R
\label{16}\\
&&\phantom{aaa}\times
\left[\rho_2-(\rho_2-\rho_1)
e^{-\frac{\Delta_2}{\lambda}}-(\rho_1-\rho)
e^{-\frac{\Delta_2+\Delta_1}{\lambda}}\right]
\times
\left[\rho_2 -
(\rho_2-\rho_1)
e^{-\frac{\Delta_2}{\lambda}}-
(\rho_1-\rho^{\prime})
e^{-\frac{\Delta_2+\Delta_1}{\lambda}}
\right].
\nonumber
\eea

Now let us consider the Yukawa-type hypothetical force
taking into account the surface distortions covering both
the sphere and the disc. One might expect that their
contribution is of prime importance when $\lambda$
is of the same order as the distortions amplitude
$A$ or even smaller. The vertical distance between
the distorted disc and lens surfaces is given by
\beq
a_d=
a-A\left[f(x+\varphi_x,y+\varphi_y)+f(x,y)\right].
\label{17}
\eeq
\nn
Here $f(x,y)$ is a function describing box-type
distortions which are the same for both bodies,
$\varphi_{x,y}$ are the  phase shifts 
of the distortions
in $x$- and $y$-coordinates between different
surfaces. They may take the values from zero till
the corresponding size of the characteristic 
distortion cell of some area $S$. Inside of this 
cell the quantity $a_d$ from \Ref{17} can have the
following values
\beq
a_d=\left\{
\begin{array}{rl}
a-2A, & for\ x,y\in S_1, \\
a-(1-\beta)A, & for\ x,y\in S_2, \\
a+2\beta A, & for\ x,y\in S_3,
\end{array} \right.
\label{18}
\eeq
\nn
when, correspondingly, there are two boxes against
each other on both surfaces, a box against a
nondistorted point or two nondistorted points
against each other. (We remind that the parameter $\beta$
characterizing the area, occupied by distortions,
was introduced in Sec.~II.) Each of the values of
$a_d$ from \Ref{18} is taken with a probability
$w_i=S_i/S$; $S\equiv S_1\cup S_2\cup S_3$.

Clearly the areas $S_i$, and, consequently, the
probabilities $w_i$, depend on $\varphi_x$,
$\varphi_y$. Averaging over all the values of
$\varphi_x$, $\varphi_y$, we get the averaged
probabilities

\beq
\langle w_1 \rangle_{\varphi_x,\varphi_y}=
\frac{\beta^2}{(1+\beta)^2},\quad  
\langle w_2 \rangle_{\varphi_x,\varphi_y}=
\frac{2\beta}{(1+\beta)^2},
\quad  
\langle w_3 \rangle_{\varphi_x,\varphi_y}=
\frac{1}{(1+\beta)^2}.
\label{19}
\eeq

Then the force between the distorted surfaces may be
represented as a linear combination of the expressions
\Ref{16} with the weights \Ref{19}

\beq
F_{Yu}^{(d)}(a)=
\frac{2\beta}{(1+\beta)^2}
F_{Yu}\left(a-(1-\beta)A\right)
+\frac{\beta^2}{(1+\beta)^2}
F_{Yu}(a-2A)+
\frac{1}{(1+\beta)^2}
F_{Yu}(a+2\beta A).
\label{20}
\eeq
\nn
Substituting \Ref{16} into Eq.\ \Ref{20} we obtain the
final result for the Yukawa-type hypothetical interaction
by accounting of distortions

\beq
F_{Yu}^{(d)}(a)=
\frac{F_{Yu}(a)}{(1+\beta)^2}\left[
\beta^2
e^{\frac{2A}{\lambda}}+
2\beta
e^{\frac{(1-\beta)A}{\lambda}}+
e^{-\frac{2\beta A}{\lambda}}\right].
\label{21}
\eeq

It is seen from \Ref{21} that for $\lambda\gg A$
the contribution of distortions is negligible.
Expanding \Ref{21} in powers of $A/\lambda$ we
find that the first nonzero correction is of 
second order
\beq
F_{Yu}^{(d)}(a)\approx F_{Yu}(a)\left(1+
\beta^2\frac{A^2}{\lambda^2}\right).
\label{22}
\eeq
\nn
So, the distortions begin to contribute when
$\lambda\lsim 80\,$nm. At $\lambda\sim 15\,$nm
they determine the hypothetical force value. Here the
Eq.\ \Ref{21} should be used in computations.

The interatomic interaction due to the exchange of
massless hypothetical particles may be described
by the power-law effective potentials
\beq
V_n=-\lambda_n N_1 N_2 \hbar c \frac{1}{r_{12}}
\left(\frac{r_0}{r_{12}}\right)^{n-1},
\label{23}
\eeq
\nn
where $\lambda_n$ are the interaction constants.
The quantity $r_0=1\,$F$=10^{-15}\m$ is introduced
to provide the proper dimensionality with different
$n$ [23].

For the force between a homogeneous sphere and 
an infinite plate one obtains with $n\geq 3$
\beq
F_n^{(sp)}(a)=-2\pi\lambda_n
\frac{\hbar c}{m_p^2(n-2)}\rho\rho^{\prime}r_0^{n-1}
\int\limits_{V_s}d^3r_1
\frac{1}{z_1^{n-2}}.
\label{24}
\eeq

These quantities depend on $a$ very slowly. For
example, integration in \Ref{24} for $n=4$ leads to the result
\beq
F_4^{(sp)}(a)=-2\pi^2\lambda_4
\frac{\hbar c}{m_p^2}\rho\rho^{\prime}r_0^{3}
\left[(R+a)\ln\frac{2R+a}{a} -2R
\right].
\label{25}
\eeq

Combining the appropriate number of expressions
\Ref{24}, \Ref{25} it is not difficult to obtain
the power-type hypothetical force with account of
metallic layers covering the sphere and the plate.
It is evident from below, however, that the
experiment [12] does not lead to any interesting
new constraints for the power-type interactions.
For this conclusion the expressions \Ref{24}, \Ref{25}
would be ample.

\section{OBTAINING OF CONSTRAINTS FOR HYPOTHETICAL
YUKAWA-TYPE INTERACTIONS}

As it was noted in Sec.~II, the theoretical
expression for the Casimir force \Ref{6} was confirmed
experimentally in [12,20] with the absolute error
$\Delta F=2\times10^{-12}\,$N. This means that the probable
hypothetical force (if any) should be constrained by the
inequality
\beq
|F^{hyp}(a)|\leq\Delta F=2\times 10^{-12}\,\mbox{N}.
\label{26}
\eeq
\nn
Here the Yukawa-type force \Ref{16}, \Ref{21} or the power-law
ones \Ref{24} or \Ref{25} can play the role of
$F^{hyp}(a)$.

The strongest constraints on the parameters of hypothetical
interactions follow from \Ref{26} for the smallest possible
value of $a=120\,$nm. Substituting \Ref{16} and \Ref{21} into
\Ref{26} with account of numerical values of all involved
quantities (see Secs.~II,\,III) we obtain constraints on the
Yukawa-type interaction following from the experiment [12]. They are
shown by the curve 1b in Fig.~2. The region below each curve
in the ($\lambda ,\alpha$)-plane is permitted by the inequality
\Ref{26} and above the curve it is prohibited. The curve 1a in
Fig.~2 indicates constraints on Yukawa-type interaction which 
would be obtained without taking into account the contribution 
of surface distortions. In this case the total hypothetical 
force is given by Eq.~\Ref{16} only. In the same figure curves 2 
and 3 correspondingly show constraints following from the old
measurements of the Casimir force [4,\,7,\,13,\,24] and of the van
der Waals force [4,\,25,\,26] between dielectrics. The constraints
given by the curves 2,\,3 were the best ones in nanometer range
up to date.

It is seen that with account of surface distortions the
experiment [12] leads to the best constraints for Yukawa-type 
interaction with a wide range of action
5.9\,nm$\leq\lambda\leq$100\,nm.
The maximal strengthening of 140 times takes place around
$\lambda =14\,$nm.
Note that when neglecting the distortion contributions to
the Yukawa force this range would be more narrow
8.7\,nm$\leq\lambda\leq$100\,nm.
In this case the maximal strengthening of constraints would be
about 65 times only. It is seen from Fig.~2 that the distortions
cease to contribute to the strength of constraints for
$\lambda\geq 25\,$nm.
The $\lambda$-range in which the constraints are strengthened in
ten times or more is
8.3\,nm$\leq\lambda\leq$32\,nm
with account of distortions. This range would be more
narrow
(11\,nm$\leq\lambda\leq$32\,nm) if the distortions are neglected.
By this it follows that the surface distortions contribute 
essentially for the Yukawa-type hypothetical force in the nanometer
range and should be taken into account in all computations of
constraints.

Now let us turn to the power-type hypothetical interactions.
Substituting, e.g., \Ref{25} into inequality \Ref{26}
one gets the constraint $\lambda_4\leq 3$. This result
is about one thousand times weaker than the constraint obtained
from the old Casimir force measurements [6]. For the power-law
interactions with smaller $n$ the results are even worse.
The reason is that the power-type forces between a plate and 
a sphere are almost independent on the distance. So, the decrease
of distance in [12] in comparison to the previous experiments 
does not lead to an increase of the force. Meanwhile, decreasing 
the radius of the sphere by one thousand times (from 10\,cm
previously to 100$\,\mu$m now) leads to the corresponding decreasing
of the hypothetical force.

The constraints, following from the experiment [12], can be strengthened 
additionally by a minor modification of experimental setup. Let us start 
with a discussion of the metallic layers covering the disc and the sphere.
It is easily seen that the contribution of their interaction to
hypothetical force determines its value almost completely. Actually,
the contribution of a layer on a sphere and a layer on a disc is given
by the combination of four expressions of the form of \Ref{14} with
appropriate parameters. For example, the force acting between two
outer layers is
\beq
F_{Yu}^{(2,2)}(a)=-4\pi^2 \alpha 
\frac{\hbar c}{m_p^2} \lambda^3
e^{-a/\lambda}\,R\rho_2^2\left(
1-e^{-\Delta_2/\lambda}\right).
\label{27}
\eeq

Performing the computations for the layers
$Au/Pd$ between theirselves we come to the conclusion that they contribute
98\% of the hypothetical force value at $\lambda=6\,$nm and 33\% 
at $\lambda=100\,$nm. In the same way the contribution of $Au/Pd - Al$
layers of both bodies is 
10\% for $\lambda=15\,$nm and
46\% for $\lambda=100\,$nm.
Finally $Al - Al$ layers contribute only
1.5\% of force at $\lambda=25\,$nm and
17\% at $\lambda=100\,$nm.
For comparison, the interaction of the sphere layers with the sapphire
disc contribute only
1\% of the hypothetical force at $\lambda=70\,$nm and
4\% at $\lambda=100\,$nm.
The interaction of the disc metallic layers with the polystyrene sphere
does not contribute to the hypothetical interaction with the required 
accuracy  not to speak of the interaction of polysterene with sapphire.

As is seen from above the main contribution to the nanometer scale
Yukawa interaction is given by the $Au/Pb$ outer layers. If to
change them for  more heavy purely $Au$  layers of the same thickness
the obtained constraints would be strengthened in 1.2 times in the
range $\lambda\leq 60\,$nm. Also, the application range of new
constraints would be a bit wider on the account of small $\lambda$.

The other possibility is to increase the thickness of $Au/Pb$ layer
till, e.g., 25\,nm (a further increasing would decrease partly the
advantage of the
good reflectivity properties of Al). This also gives the possibility
to strengthen constraints for Yukawa interaction in 1.2 times in
the range $\lambda\leq 40\,$nm.

If to combine both suggestions, i.e., to use purely $Au$ layers of
25\,nm thickness, the obtained constraints would be in 1.4 times stronger.

More radical strengthening can be obtained by the use of a larger sphere.
Due to the linear dependence of the hypothetical force \Ref{16} on a
sphere radius $R$ the increasing of it in, e.g., 3 times would lead to
the same strengthening of constraints (in this case metallic layers should
cover only the top of the sphere, nearest to a disc, not to make the sphere
too heavy; also a semisphere may be used with the same success).
It seems to be possible also to decrease the absolute error of force
measurements in 2 times, i.e., till $10^{-12}\,$N. This would strengthen
constraints in 2 times simultaneously.

As a result of all these suggestions the obtained constraints could be
strengthened in 8.4 times without any principal change of the setup.
The maximal strengthening of the known up date constraints for
Yukawa-type interaction in nanometer scale would be about 1200 times.

\section{CONCLUSION AND DISCUSSION}

As is seen from above the new measurement of
the Casimir force using an atomic force microscope gives
the possibility to strengthen constraints for hypothetical
Yukawa-type interaction in nanometer scale. The strengthening
takes place in the range
5.9\,nm$\leq\lambda\leq$100\,nm.
The maximal strengthening in 140 times occurs around
$\lambda=$14\,nm.
It is interesting to compare this result with the constraints
for Yukawa interaction following from the other recent experiment
on measuring of the Casimir force [8]. According to the results
of Ref.~[11], where these constraints were obtained, the
strengthening takes place in the range
220\,nm$\leq\lambda\leq 1.6\times 10^5\,$nm.
The new constraints surpass the old ones following from the
measurements of the Casimir force between dielectrics up to
a factor of 30. Thus, both recent experiments give  complementary
results which are valid in different regions. At the same time
there is a wide gap for
100\,nm$<\lambda < 220$\,nm
where the former constraints are valid [4,7] obtained from the
Casimir force measurements between dielectrics [13,24]. It should
be covered by future experiments on measuring the Casimir force
(see, e.g., a proposal to measure the Casimir force using the
suspended Michelson interferometer developed for gravitational
waves detection [27]).

There is a tendency of widening the range of $\lambda$ for which
the Casimir effect leads to the strongest constraints on Yukawa
hypothetical interactions. According to [4,7] the former measurements 
of the Casimir force were the source of strongest constraints in the
range
$10^{-8}\m\leq\lambda\leq 10^{-4}\m$.
At present, combining the results following from the experiments [8,12],
we get the strongest constraints in a wider range
$5.9\times 10^{-9}\m\leq\lambda\leq 1.6\times 10^{-4}\m$.
This means that the Casimir effect not only succeeded in obtaining
stronger constraints for hypothetical interaction but also
successfully compets with the measurements of van der Waals forces
(in $\lambda <10^{-8}\m$ range) and with Cavendish-type experiments
(for $\lambda >10^{-4}\m$).
In this sense the experiments on the Casimir force measuring suggest
a good supplement to the other experimental investigations giving
stronger constraints for the hypothetical interactions (see, e.g., 
[28] where the new constraints were obtained for 
$\lambda\approx 0.5\m$ from the short-range test of equivalence
principle or [29] where the strengthening for extremely large
$\lambda$ was achieved from the satellite measurement of the 
Earth's magnetic field).

As it was discussed in the preceeding section the constraints, obtained
in this paper, could be strengthened by a factor of about 8 due to
some modifications of experimental setup. As a result the maximal
strengthening in nanometer scale may achieve of about 1000 times.
According to the results of Ref.~[11] the other experiment [8] on
the Casimir force measuring also could lead to much stronger
constraints being modified in appropriate way. This would give the 
possibility to constrain masses of such hypothetical elementary
particles as graviphoton and dilaton by the use of the Casimir force
measurements (see Ref.~[30] which contains some  experimental
evidence for the existence of graviphoton obtained from
geophysical data).

\section*{ACKNOWLEDGMENTS}
%\acknowledgments

The authors are especially grateful to U.~Mohideen for important
additional information concerning his and A.~Roy's experiment and
numerous helpful discussions about the accuracy of force
measurements and surface distortions. G.L.K.\ and V.M.M.\ are indebted 
to the Institute of Theoretical Physics of Leipzig University,
where this work was performed, for kind hospitality. G.L.K.\ was
supported by Saxonian Ministry for Science and Fine Arts. V.M.M.\ was
supported by Graduate College on Quantum Field Theory at Leipzig
University.

%%%%%%%%%%%%%%%%%%%%%%%%%%%%%%%%%%%%%%%%%%%%%%%%%%%%%%%
%\newpage

\begin{figure}[h]
\epsfxsize=10cm\centerline{\epsffile{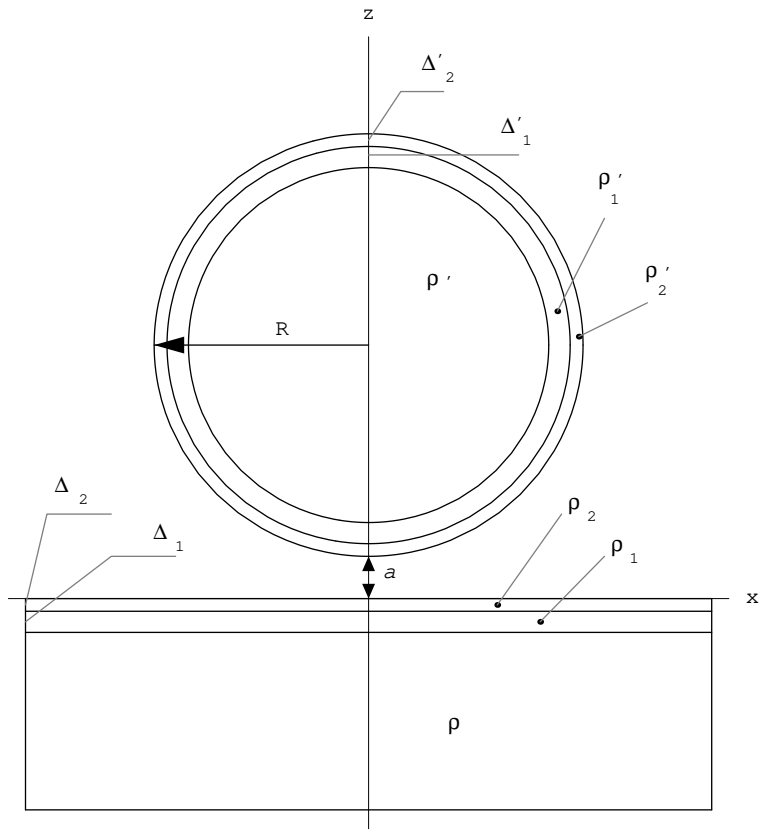} }
\caption{Configuration of a sphere of radius
$R$ and density $\rho^{\prime}$
 above a disc of density $\rho$
spaced at a distance $a$. The thicknesses and densities of $Al$
and $Au/Pd$ layers on the disc are $\Delta_1,\,\Delta_2,$ 
$\rho_1, \,\rho_2$ and on the
sphere are $\Delta_1^{\prime},\,\Delta_2^{\prime},$
$\rho_1^{\prime},\,\rho_2^{\prime}$,  respectively.
}
\end{figure}
\begin{figure}[h]
\epsfxsize=10cm\centerline{\epsffile{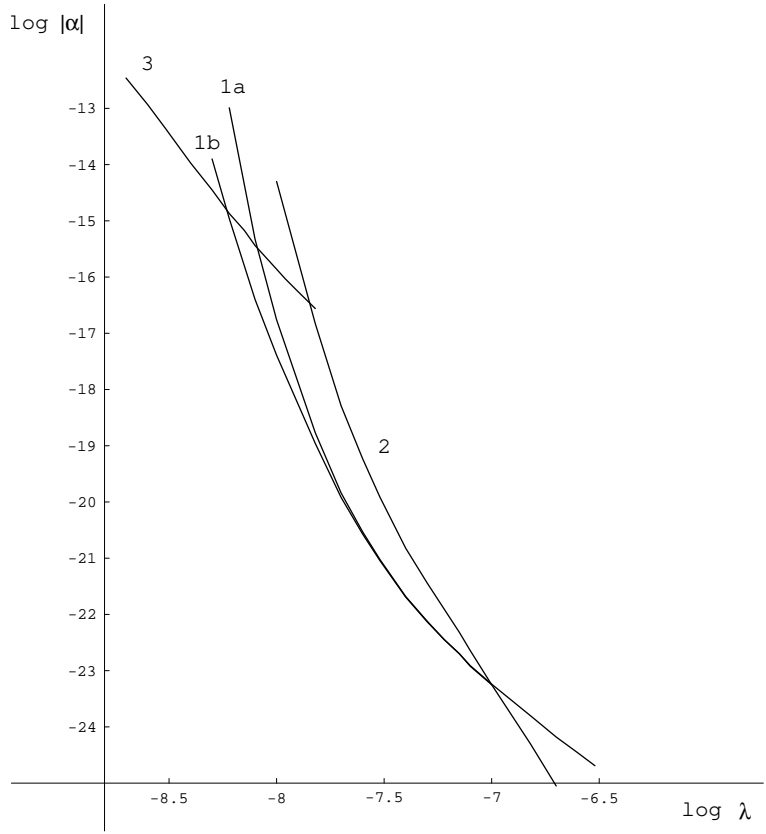} }
\caption{Constraints for the constants of
hypothetical Yukawa-type interactions following from different
experiments. Curves 1a, 1b are obtained in this paper. They
follow from the new measurement of the Casimir force [12] using
an atomic force microscope (a --- without account of surface
distortions, b - with account of distortions). Curve 2 follows
from the measurement of the Casimir force between dielectrics
[4,7,13,24], curve 3 results from the van der Waals force between 
crossed cylinders [25,26].
}
\end{figure}

\end{document}